\documentclass[prl,twocolumn,amsmath,amssymb,groupedaddress,superscriptaddress]{revtex4}
\usepackage{graphicx}
\usepackage{color}

\begin{document}
\title{Fragment Percolating and Many Body Localization Transition}
\author{Xiaolong \surname{Deng}}
\affiliation{Institut f\"ur Theoretische Physik, Leibniz Universit\"at Hannover, Appelstr. 2, 30167 Hannover, Germany}

\begin{abstract}
Elements of eigenvectors obtained by exact diagonalization can be considered as two dimensional lattice sites, in which dynamics of a given initial state is seen as a percolating procedure on the lattice sites. Then one can use the percolating procedure to generate a series of fragments of eigenvectors. Combining the fraction ratio and entanglement entropy dynamics of the generated fragment we examine the many-body localization transition in a disordered hard-core bosonic model. It helps us to locate roughly the transition point at the critical disorder strength $W_c\approx 12\sim13$. The scaling collapse of the fraction ratios gives us the localization length exponent $\nu\approx2.0$.

\end{abstract}
\date{\today}
\maketitle



\paragraph{Introduction.--}
Interactions thermalize all of parts of a system\cite{Deutsch1991,Srednicki1994}, while disorder localize non-interacting particles\cite{Anderson1958,Mirlin2008}. The interplay of disorder and interactions in quantum many-body systems can lead to fascinating behaviors. One of prominent examples is many-body localization (MBL)\cite{Basko2006, Gornyi2005}, in which quenched disorder prevents thermalization of interacting systems. MBL provides a generic mechanism of ergodicity breaking in quantum systems, and has attracted a huge attention in recent years~\cite{Nandshikore2015,Alet2018,Abanin2019}, including breakthrough experiments~\cite{Schreiber2015,Choi2016,Bordia2016,Smith2016,Lueschen2017,Lukin2019,Rispoli2019}.

MBL systems are characterized by extensive number of local integral motions (LIOMs)\cite{Serbyn2013b,Huse2014}. The LIOMs forming a complete set can completely determine the eigenstates. In the deep MBL phase two eigenstates with the same total energy and particle number generally have different values of LIOM. Thus, each eigenstate itself will form a fragmentation. Extensive fragmentations break down the thermalization and lead to non-ergodicity. Very recently, the idea of Hilbert space fragmentation has been explored extensively not only in disorder-free systems \cite{Khemani2019,Sala2019,Exp_fragmentation2020}, but also for the MBL problems\cite{LoganChalker2018,Pietracaprina2019,DeTomasi2019,Tarzia2020,Prelovsek2020,Krylov2020}. In Ref. \cite{LoganChalker2018} the authors argue that the MBL transition can be represented by a classical percolation model constructed in a graph of Fock space. The percolation is decided by the competition between tunneling and energy separation of Fock basis states. For the disordered Ising model they get the scaling exponent $\nu=2$, which is consistent with the Harris bound\cite{Harris1974,CCFS1986}. Based on a similar criterion as in \cite{LoganChalker2018}, Ref. \cite{Pietracaprina2019} focus on a decimation scheme through eliminating the irrelevant parts of Hilbert space. In Ref. \cite{Prelovsek2020} the authors use rate equations to emphasizes the percolation aspect of the MBL also. In their work they cluster the eigenstates with the most largest overlaps into a fragment, and give the critical disorder $W_c\approx 8$ for the random Heisenberg model.



\begin{figure}[t]
	\includegraphics[width=1\columnwidth]{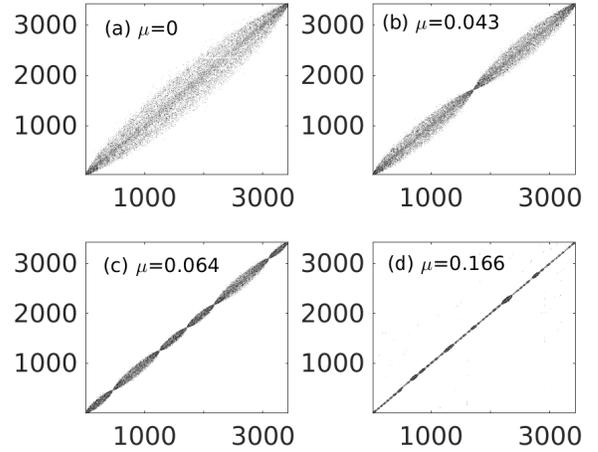}
	\caption{The percolating procedure generates fragment structures of eigenstates under different strengths of cutoffs $\mu$: (a) $\mu=0$, the original structure with ascending energy indices; (b) $\mu=0.043$; (c) $\mu=0.064$; and (d) $\mu=0.1664$. 
		The $x$ and $y$ coordinates correspond to the indices of eigen states and Fock states, respectively. Inside the fragment the indices are sorted in ascending energy order. The calculation is done for $L=14$, $N=7$, $V=1$ and $W=15$ with open boundary condition and a single disorder.}
	\label{fig:1}
\end{figure}


 Motivated by Refs \cite{LoganChalker2018,Pietracaprina2019,Prelovsek2020}, we explore a bosonic MBL problem, which is interesting for a large variety of physical problems\cite{Yan2013,Richerme2014,Jurcevic2014,Baier2016,Leseleuc2019}. The fragment structure of the Hamiltonian eigenstates can be self-consistently selected through a percolating procedure\cite{ShklovskiiEfros}. Tracking the fragment generated by a given initial Fock state (i.e., track the area that the state is percolating), we examine the MBL transition. As the MBL transition is a kind of dynamical behavior\cite{Abanin2019}, the static quantities, such as level spacing and inverse participation ratio, might be difficult to capture its dynamical feature. The fragments are dynamically generated in the percolating procedure. The corresponding fraction ratios to the whole Hilbert space and dynamics of entanglement entropy are naturally of dynamical characteristics, which allow us to locate the MBL transition. We get numerically the critical disorder strength $W_c\approx 12\sim13$, which is much larger than the ones obtained by the level spacing\cite{footnote1}. More interesting, the scaling of fraction ratios satisfies the Harris criterion\cite{Harris1974,CCFS1986}. Our data gives $\nu\approx2.0$, where $\nu$ is the localization length exponent.

\paragraph{Model.--}
We consider a 1D hard-core extended Bose-Hubbard model. The Hamiltonian is:
\begin{eqnarray}
\hat H = \sum_{i}-J(\hat b_i^\dag\hat b_{i+1}+\mathrm{h.c.})  + V\hat n_i \hat n_{i+1} + \epsilon_i \hat n_i,
\label{eq:H}
\end{eqnarray}
where $\hat b_i$ are bosonic operators at site $i$~($(\hat b_i^\dag)^2=0$), $\hat n_i=\hat b^{\dagger}_i \hat b_i$,  
$J=1$ and $V$ are, respectively, the hopping amplitude and interaction strength, and $\epsilon_i\in[-W,W]$ is a uniformly distributed random energy. Eq.~\eqref{eq:H} is equivalent to a spin-$1/2$ random Heisenberg Hamiltonian, and has been studied intensively as a standard MBL model\cite{Luitz2015}. In the Fock basis $\{|j\rangle\}$, Eq. \eqref{eq:H} can be written as $
\hat{H} = \sum^{\mathcal{N}}_{j=1} \langle j|\sum_i n_i n_{i+1} + \epsilon_i n_i| j\rangle|j\rangle\langle j| + J\sum_{j,k}|j\rangle\langle k|,
$
Then it is considered as a single-particle Anderson problem in a high-dimensional lattice, where the first term is the effective on-site disorder, and the second is the hopping between Fock basis.

\begin{figure} [t]
	\includegraphics[width=\columnwidth]{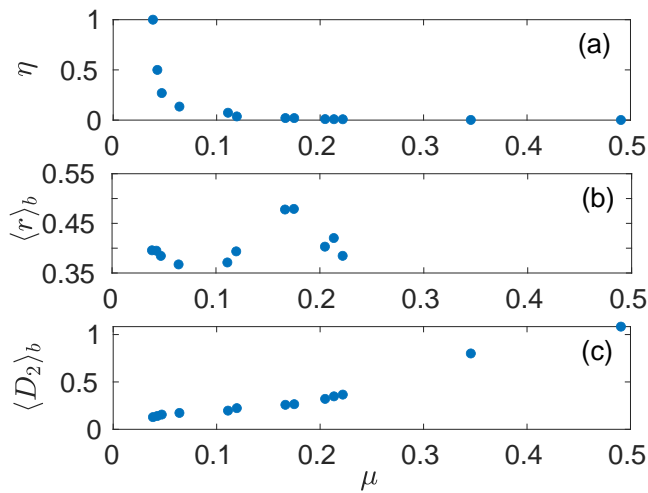}
	\caption{ Relevant quantities (a) the fragment ratio of $N_{b}/\mathcal{N}$, (b) the level spacing $\langle{r}\rangle_b$, and (c) the generalized fractal dimensions $\langle{D_2}\rangle_b$ in the fragment generated from the N\'eel state are calculated under different cutoffs. The parameters are the same as in Fig. \ref{fig:1}.}
	\label{fig:block}
\end{figure}

\paragraph{Percolating procedure.--}
By exact diagonalization of Eq. \ref{eq:H} one gets eigenenergies $E_{\alpha}$ and eigenvectors $\psi_{\alpha}(j)$. The eigenvetors $\psi_{\alpha}(j)$ form a matrix with the column eigenbasis $\{|\alpha\rangle\}$ and the row Fock basis $\{|j\rangle\}$. 

Now based on $\psi_{\alpha}(j)$ let us formulate a procedure in percolation scheme\cite{ShklovskiiEfros}. Consider a two dimensional lattice formed by  $x$-axis (eigen basis) and $y$-axis (Fock basis). The lattice sites are the elements of eigenvectors. We assume that the sites can be unblocked (active) or blocked (inactive), which are decided by the amplitudes of elements. Bonds between sites are hyperlinked either in $x$ or in $y$ direction, and unblocked. The blocked sites do not permit flow of liquid in either $x$ or $y$ direction, but the unblocked sites permit the flow of liquid in both directions. In other words, blocked sites cannot be wet, nor can wet others, while the unblocked sites are wet, and will instantly wet other unblocked sites. A given random unblocked sites through this procedure may create two possibilities: either a finite or an infinite number of unblocked sites will be wet\cite{footnote2}. A finite number of unblocked sites corresponds to a fragment of Hilbert space, while the infinite number of unblocked sites corresponds to the whole Hilbert space. Clearly the outcome will depend on the fraction of unblocked sites. Here it will be decided by the disorder strength $W$. The random positions of blocked and unblocked sites are also important. For a given disorder sample, the distribution of blocked and unblocked sites are fixed and remains constant during the percolation. In practice in our work, the fraction of unblocked and blocked sites will be decided by a cutoff $\mu$, which will be discussed later.

\begin{figure} [t]
	\includegraphics[width=\columnwidth]{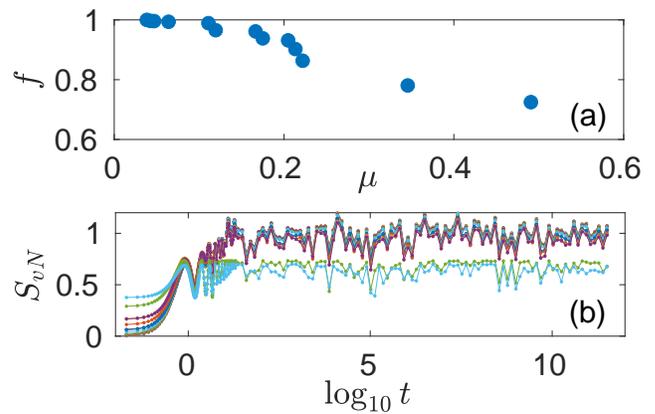}
	\caption{The fidelity $f$ (a) and the entropy dynamics (b) in the generated fragments of the N\'eel state. The different colors represent different cutoffs $\mu$. When $0\leq\mu\leq 0.2219$ the entropies are almost the same. The parameters are the same as in Fig. \ref{fig:1}.}
	\label{fig:block:entropy}
\end{figure}

We start from an initial Fock state. In the eigenbasis it means a cluster of random-distributed unblocked sites in a 2D lattice. We choose a cutoff $\mu$ to numerically decide unblocked ($>\mu$) and blocked ($<\mu$) sites. From the inverse participation ratio we know that the delocalized states have $IPR = \sum_j |\psi_{\alpha}(j)|^4 \approx \mathcal{N}|\psi_{\alpha}(j)|^4\propto 1/\mathcal{N}$, that is, $|\psi_{\alpha}(j)|\propto 1/\sqrt{\mathcal{N}}$. It will be natural to start a cutoff from $\mu = 1/\sqrt{\mathcal{N}}$, and increase it gradually. In a long time limit the percolation will be in an equilibrium. All of the connected unblocked sites may form a close subsystem. If the Hilbert space fragmentation exists here, the norm of the generated fragment should be equivalent to $1$. So, the norm $f=\sum_{j,\alpha\in{\mathrm{block}}}|\psi_{\alpha}(j)|^2$ can be used to measure the fidelity of the fragment in this block structure close to a Hilbert space fragmentation. On the other hand, we are interested in the entanglement entropy dynamics of the fragment. For an isolated finite system, observing the unitary dynamics of an initial Fock state in the whole Hilbert space should be equivalent to observe its dynamics in its Hilbert space fragmentation. Therefore, we can define a kind of fidelity of the  entropy dynamics in the fragment comparing to the whole Hilbert space. Suppose $S_{vN}(t)$ is the entropy dynamics in the whole Hilbert space and $S^b_{vN}(t)$ is the entropy dynamics in the generated fragment for a given cutoff. Then we introduce entropy ratios $\chi_{s, k}$ between $S^b_{vN}(t)$ and $S_{vN}(t)$. When the entropy is saturated in the long time limit, we measure the entrop itself $\chi_{s}=S^b_{vN}(t\to\infty)/S_{vN}(t\to\infty)$.
However, before the saturation, when the entropy is still increasing with time, we measure its slope $\chi_{k}=k^b_{vN}/k_{vN}$.
where the slope k is extracted from either $\ln{S_{vN}(t)}\propto k \ln{t}$ for the delocalized state or $S_{vN}(t)\propto k \ln{t}$ for the MBL state\cite{Znidaric2008,Bardarson2012,Serbyn2013a}. These relative-error-like quantities will measure the fidelity of entropy dynamics in the fragment. 

In connection with the percolation, let us define the fraction ratio $\eta$ of the size of generated fragment $N_b$ to the size of whole Hilbert space $\mathcal{N}$: $\eta=\frac{N_b}{\mathcal{N}}$. $\eta$ is equivalent to the probability of a random unblocked site wetting $N_b$ other unblocked sites.
In infinite lattice when $\eta=0$, the blocked sites prevent the liquid spreading far from the initial site. On the contrary, as $\eta$ approaches $1$, any unblocked site will result in all sites wet.
In disorder ensemble there are the typical fraction ratio $\eta_{typ}=\exp(\langle\ln{N_b}\rangle)/{\mathcal{N}}$, and the averaged fraction ratio $\eta_{avg}=\langle{N_b}\rangle/{\mathcal{N}}$\cite{LoganChalker2018}. Generally $\eta_{typ}$ and $\eta_{avg}$ give similar behaviors. However, in our calculations $\eta_{typ}$ and $\eta_{avg}$ give peculiar feature in the ergodic phase and the MBL phase, respectively.

\begin{figure} [t]
	\includegraphics[width=\columnwidth]{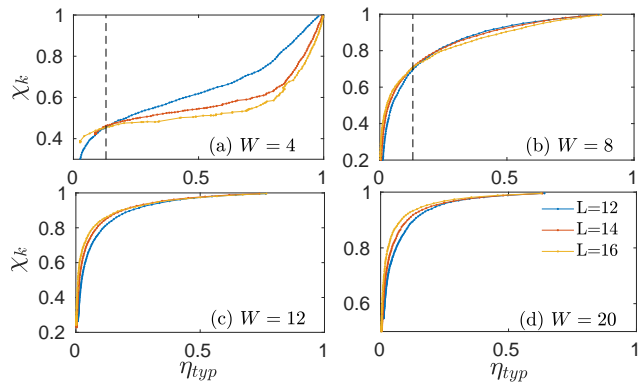}
	\caption{ The averaged entropy ratio $\chi_k$ as a function of the typical fraction ratio $\eta_{typ}$ for various systems sizes $L$ at different disorder strengths: (a) $W=4$; (b) $W=8$; (c) $W=12$; and (d) $W=20$. (a) and (b) indicate a large amount of fractions $\eta_{typ}\to\infty$ when $L\to\infty$. Starting from $W\approx12$ all of fractions $\eta_{typ}\to0$ with $L\to\infty$. The disorder samples are $2000,1000,500$ for $L=12,14,16$, respectively.
	}
	\label{fig:typ}
\end{figure}
\begin{figure} [t]
	\includegraphics[width=\columnwidth]{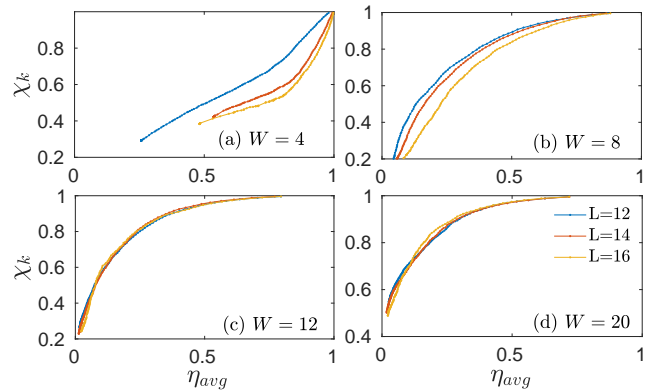}
	\caption{ The averaged entropy ratio $\chi_k$ as a function of the averaged fraction ratio $\eta_{avg}$ for various systems sizes $L$ at different disorder strengths: (a) $W=4$; (b) $W=8$; (c) $W=12$; and (d) $W=20$. (a) and (b) indicate all of fractions $\eta_{avg}\to\infty$ when $L\to\infty$. Starting from $W=12$ (c) all of fractions $\eta_{avg}$ do not depend on the system sizes. At $W=20$ (d) we have not seen the dependence. The disorder samples are $2000,1000,500$ for $L=12,14,16$, respectively.}
	\label{fig:avg}
\end{figure}

\paragraph{Fragment in a single disorder.--}
Combining $f$ , $\chi_{k,s}$ and $\eta$ we will discuss the fragment generated from an initial state. Without loss of generality, we choose the N\'eel state. We firstly choose the phase at a strong disoder $W=15$ to demonstrate our method. Fig. \ref{fig:1} shows that the percolating procedure under different cutoffs generates different fragment structures for the eigenvectors in a single disorder. The $x$-axis and $y-$axis correspond to the indices of eigen and Fock basis, respectively. At $\mu=0$ (Fig. \ref{fig:1} (a)), there is only a big block structure with eigen and Fock basis in the ascending-energy ascending order. When the cutoff is increased to some point, two fragments start to appear, and they keep the two-fragment structure until $\mu=0.043$, see Fig. \ref{fig:1} (b). Then another different structure with more fragments starts to appear, and keep itself for a range of $\mu$. With increasing the cutoff further, more and more fragments appear, see Fig. \ref{fig:1} (c) and (d). We emphasize that the same fragment structure can be kept for a range of $\mu$. In Fig. \ref{fig:1} the structures are shown at the maximal value of $\mu$ which still generate the same fragment structure.

We investigate the relevant quantities under different cutoffs. Fig. \ref{fig:block} (a) shows that the fraction ratio changes with different cutoffs. As the eigen levels in the fragment are sorted, we can calculate the level spacing $\langle{r}\rangle_b=\langle\frac{\min(E_{\alpha+1}-E_{\alpha},E_{\alpha}-E_{\alpha-1})}{\max(E_{\alpha+1}-E_{\alpha},E_{\alpha}-E_{\alpha-1})}\rangle_b$\cite{Huse2007} and the generalized fractal dimensions $\langle{D_2}\rangle_b=-\ln\langle{IPR}\rangle_b/\ln{N_{b}}$\cite{Mirlin2008}, see Fig. \ref{fig:block} (b) and (c). 

Through checking the fidelity of the series of fragments in Fig. \ref{fig:block:entropy} (a) we know that our percolating procedure works very well. Comparing Fig. \ref{fig:block:entropy} (a) and (b) we find that even with $f=0.85$ the fragment entropy dynamics is still close to the dynamics in the whole Hilbert space. When $\mu<0.23$ all of generated fragments give almost the same dynamics. Therefore, one may expect an as-small-as possible fragment but it can still capture the physics in the whole Hilbert space. 

\paragraph{MBL transition.--}

We have seen the cutoff triggers the percolation in our procedure. However, various disorder samples may induce different cutoffs, different percolated fractions, and different fidelities. This means we cannot simply fix either the cutoff or the fidelity to make average over disorder ensemble. For individual disorder we have to scan the cutoff from $0$ to the maximal value (supporting the minimum of the fragment fraction). Before making average we take into account the relative errors of saturated entropy $1-\chi_s$. For each sample we account only the smallest fraction within a given relative error $1-\chi_s$, and then make average over samples. In this way, we get the typical (averaged) fraction ratio $\eta_{typ}$($\eta_{avg}$). The relation between $\chi_{k,s}$ and $\eta_{typ,avg}$ will help us to roughly locate the transition.

In Fig. \ref{fig:typ} we plot the averaged entropy ratio as a function of the typical fraction ratio for various system sizes $L$. Fig. \ref{fig:typ} (a) and (b) show that both at small disorder strength $W=4$ and at finite disorder strength $W=8$, $\eta_{typ}$ increases with system sizes $L$. This indicates that $\eta_{typ}\to 1$ when $L\to\infty$, although the speed at $W=8$ might be much slower than the one at $W=4$. It is interesting to note that when $\eta_{typ}<0.1$ (see the left sides of the dashed lines in Fig. \ref{fig:typ} (a) and (b)) the fragment tends to be localized (i.e. $\eta_{typ}\to0$ at $L\to\infty$), and with a high entropy ratio for $W=8$. This unconventional fragment is only seen in the regime $W\in[4,8]$ of $\eta_{typ}$. The averaged fraction ratio $\eta_{avg}$, however, gives us clearly the tendency to the delocalization for all of fragments at $W=4,8$, see Fig. \ref{fig:avg} (a) and (b).

With the disorder strength increasing, the localization effect becomes stronger. From the curves of $\chi_k-\eta_{typ}$ and $\chi_k-\eta_{avg}$ we can roughly determine the MBL transition point. At the transition, the unconventional regime starts to extend in the whole $\eta_{typ}$. All of $\eta_{typ}\to0$ when $L\to\infty$, see Fig. \ref{fig:typ} (c). In this sense it indicates that at the transition point may be a localized phase. Comparing to $\eta_{typ}$, Fig. \ref{fig:avg} (c) shows that $\eta_{avg}$ does not depend on the system sizes at the transition point. This indicates that there might be a universal distribution of $\eta_{avg}$ at the critical $W_c$ (see Fig. \ref{fig:avg} (c)). Increase further the disorder, however, $\eta_{avg}$ does not show the clear dependence on the system sizes, at least at $W=20$ (Fig. \ref{fig:avg} (d)). This is consistent with the observation in Ref. \cite{Pietracaprina2019}.

\begin{figure} [t]
	\includegraphics[width=\columnwidth]{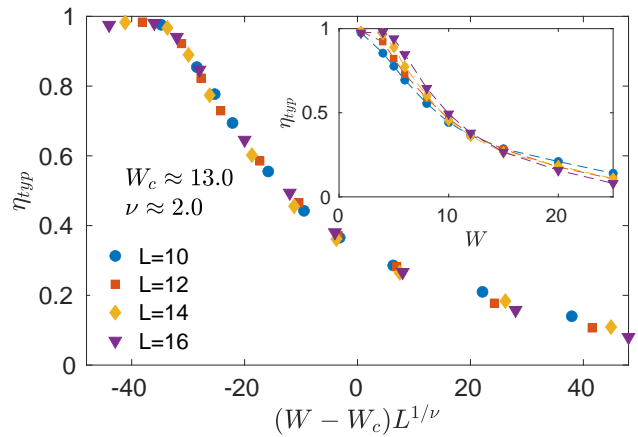}
	\caption{ The typical fraction ratio at $\chi_s=0.98$ as a function of disorder strength for various systems sizes $L$. The main panel shows the scaling collapse with $W_c\approx13.0$ and $\nu\approx2.0$. The inset are the raw data before collapsing. The disorder samples are $5000,2000,1000,500$ for $L=10,12,14,16$, respectively.
	}
	\label{fig:block:mbl}
\end{figure}

We have obtained the critical disorder $W_c$ from the growth of the entropy dynamics $\chi_k$. Now we would like from the other side to check the transition using the saturated entropy dynamics $\chi_s$. For a given $\chi_s$ we do the scaling collapse of $\eta_{typ}$ for various systems, by the means of the universal function $(W-W_c)L^{1/\nu}$, where $\nu$ is the localization length exponent. It is known that a disorder-driven phase transition in $d$ dimensional space satisfies the Harris bound $\nu\geq 2/d$\cite{Harris1974,CCFS1986}. In MBL transitions a similar bound is also derived\cite{Chandran2015,Khemani2017}. In our scaling collapse we choose the values of $\chi_s$ close to $1$, for instance $\chi_s=0.98$. It is very interesting that our data yields $W_c\approx 13$ and $\nu\approx 2.0$, see Fig. \ref{fig:block:mbl}. Our result on the scaling exponent is consistent with Ref. \cite{LoganChalker2018}.

\paragraph{Conclusions and outlook.--}
Given an initial Fock state we can generate the fragments of the eigenvectors of a Hamiltonian by the means of a percolating procedure. Using this method we have investigated the hard-core extended Bose-Hubbard model with on-site disorder. Combing the fraction and entropy ratios of the fragments we have shown that the MBL transition may take place at about $W_c\approx 12\sim 13$, and the fraction ratios have a scaling collapse with $\nu\approx2$.  

This percolating procedure can also be applied to power-law interacting/hopping lattice models\cite{Burin2006,Yan2013,Yao2013,Burin2015a,Burin2015b,Deng2020}. In future it will be interesting to study whether the generated fragment depends on the power-law exponent.


\paragraph{Acknowledgements.--}We are grateful to Luis Santos and Wei-Han~Li for collaboration on related  work and for very useful discussions. We acknowledge the support of the German Science Foundation~(DFG) (SA 1031/11, SFB 1227, and Excellence Cluster QuantumFrontiers).



\newpage


\begin{thebibliography}{99}
\bibitem{footnote2} It is generally accepted that there can be no more than one infinite cluster in a lattice\cite{ShklovskiiEfros}.
\end{thebibliography}
\end{document}